\documentclass[aps,pre,reprint,superscriptaddress,showpacs]{revtex4-1}
\usepackage{graphicx}
\usepackage{amsmath}
\usepackage{amssymb}
\usepackage{psfrag}

\newcommand {\e} {\varepsilon}
\def\W{\Omega}

\def\w{\omega}

\begin{document}
\title{Marginal chimera state at cross-frequency locking of pulse-coupled neural networks}
\author{M. I. Bolotov}
\affiliation{Department of Control Theory, Nizhni Novgorod State University,
Gagarin Av. 23, 606950, Nizhni Novgorod, Russia}
\author{G. V. Osipov}
\affiliation{Department of Control Theory, Nizhni Novgorod State University,
Gagarin Av. 23, 606950, Nizhni Novgorod, Russia}
\author{A. Pikovsky}
\affiliation{Institute for Physics and Astronomy, 
University of Potsdam, Karl-Liebknecht-Str. 24/25, 14476 Potsdam-Golm, Germany}
\affiliation{Department of Control Theory, Nizhni Novgorod State University,
Gagarin Av. 23, 606950, Nizhni Novgorod, Russia}

\begin{abstract}
We consider two coupled populations of leaky integrate-and-fire neurons. Depending
on the coupling strength, mean fields generated by these populations can have incommensurate
frequencies, or become frequency locked. In the observed 2:1 locking state
of the mean fields, individual neurons
in one population are asynchronous with the mean fields, while in another populations
they have the same frequency as the mean field. These synchronous neurons form a 
chimera state,
where part of them build a fully synchronized cluster, while other remain scattered.
We explain this chimera as a marginal one, caused by a self-organized neutral
dynamics of the effective circle map. 
\end{abstract}
\date{\today}
\date{\today}

\maketitle
\section{Introduction}
Studies of the dynamics of globally coupled populations of oscillators, 
pioneered more than 40 years ago by Winfree and 
Kuramoto~\cite{Winfree-67,*Kuramoto-75}, are in the focus of current research
due to numerous applications in diverse fields from physics to neuroscience,
but also due to striking effects like synchronization, collective chaos, and 
chimera states~\cite{Pikovsky-Rosenblum-15}. 

While typically ensembles of identical
oscillators either fully synchronize or desynchronize, depending on whether the coupling 
is attractive or not, there are situations where oscillators produce a macroscopic mean field
without full synchrony, such a regime is called partial 
synchronization~\cite{vanVreeswijk-96,Rosenblum-Pikovsky-07} (see
Ref.~\cite{Temirbayev_etal-12} for its experimental observation). Remarkably,
partial synchronization can be explained within a simplest setup of one-dimensional 
oscillators, described either by their phase dynamics~\cite{Rosenblum-Pikovsky-07},
or as integrate-and-fire units~\cite{vanVreeswijk-96}.
Another way of breaking full synchrony is
formation of a chimera state, where part of a population of identical oscillators
synchronize and form a cluster, while other remain asynchronous. This regime requires self-induced bistability,
so far such regimes have not been observed in one-dimensional phase models, but in  
oscillators described by high-dimensional 
equations~\cite{Schmidt_etal-14,*Sethia-Sen-14,*Yeldesbay-Pikovsky-Rosenblum-14}. 

In this paper we study mutual coupling of two populations of partially synchronous 
integrate-and-fire
oscillators, and find a surprising chimera state in this system: at mutual 2:1 
locking of two macroscopic
mean fields, elements of one ensemble form a cluster and a marginally 
stable scattered group.
This means that nontrivial chimeras can happen in populations of one-dimensional
units. Below we will analyze how the mean field dynamics, yielding effective bistability
in such ensembles, appears in a self-consistent way. 

Before proceeding to description of the model, we mention that
mutual influence of two or several populations of oscillators, generating macroscopic mean fields at
significantly different frequencies,   has been studied recently in the context of
phase dynamics of Kuramoto model type~\cite{Komarov-Pikovsky-11,*Komarov-Pikovsky-13,*Komarov-Pikovsky-15a}.
Here we extend these studies to realistic neural models of integrate-and-fire neurons, which is applicable
for explanation of cross-relations between brain waves. Indeed, in brain one
observes macroscopic oscillations in different frequency ranges~\cite{Buzsaki-06}, and these 
realtively regular mean fields
are not related to exact synchrony of individual neurons, but 
rather to a temporal organization of their firing events.
Thus, the model of partial synchronization we consider below, appears 
more adequate for the neural dynamics than the Kuramoto model of phase oscillators.

\section{The Model}
Our consideration of two neural populations is based on the model
of globally coupled leaky integrate-and-fire oscillations, a prototypical
example of generation of nontrivial mean fields due to partial synchronization~\cite{vanVreeswijk-96}.
One such population has been thoroughly studied in 
Refs.~\cite{vanVreeswijk-96,Mohanty-Politi-06,Zillmer_etal-07}. 
The potential $x_k$
of each neuron (here $1\leq k\leq N$, $N$ is the size of the population) 
is described by the following equation
\begin{equation}
\dot{x}_k=a-x_k+g E\;,
\label{eq:1x}
\end{equation}
where the mean field $E$ is composed by contributions from all the neurons
\begin{equation}
\ddot{E}+2\alpha \dot{E}+\alpha^2 E=\frac{\alpha^2}{N}\sum\delta(t-t_{k,j})\;.
\label{eq:1E}
\end{equation}
The potential of neuron $k$ grows from $0$ to the threshold value $1$, governed
by the suprathreshold input current $a>1$ and the excitatory action from the field $E$
(Eq.~\eqref{eq:1x}).
When this potential reaches $1$ at a time instant $t_{k,n}$, 
the neuron fires, contributing a delta-function pulse 
to the field $E$ (Eq.~\eqref{eq:1E}), and is reset to $0$. 
Linear equation~\eqref{eq:1E} describes thus a sequences of so-called $\alpha$-pulses
created by the spiking neurons. As the analysis of system (\ref{eq:1x},\ref{eq:1E})
has shown (see Refs.~\cite{vanVreeswijk-96,Mohanty-Politi-06,Zillmer_etal-07} for details), 
for large values of the coupling parameter $g$ (at given $\alpha$)
the neurons are desynchronized:
the time intervals $t_{k,n}-t_{k-1,n}$ between successive firing events of two neurons 
are constant (do not depend on $k$), and the field $E$ is nearly a constant (with small 
variations $\sim N^{-1}$).  At some critical value of coupling $g$ this regime 
becomes unstable, and the neurons start to form a group with a smaller interval between 
firings, as a result the mean field $E(t)$ demonstrate macroscopic nearly 
periodic variations~\cite{vanVreeswijk-96,Mohanty-Politi-06}. This state is called partial synchronization, because 
neurons never synchronize fully (never fire simultaneously). In fact, in this state
the dynamics is, strictly speaking, quasiperiodic, because frequency of spiking of
a neuron $\w$ (it this the same for all neurons because they are identical) is incommensurate
to the frequency of the macroscopic mean field oscillations $\W$.

In this paper we consider two interacting populations of neurons of the described type.
Contrary to Ref.~\cite{Olmi-Politi-Torcini-10}, where two identical populations have been considered, we  
study two different populations (while inside each population all the neurons are 
identical), therefore we use slightly changed notations: membrane potentials of the 
neurons in the populations will be denoted as $x_k$ and $y_k$, while the mean fields generated
by them as $X$ and $Y$, respectively. Interaction is due to a mixture of the mean fields:
on neurons $x$ acts the field $E_x=(1-\e)X+\e Y$, while on neurons $y$ acts the field 
$E_y=(1-\e)Y+\e X$. Parameter $\e$ describes the coupling between populations. Equivalently,
one can consider this setup as one with two acting mean fields $(E_x,E_y)$, which are fed 
by spikes from two populations: each spike of a neuron $x$ contributes a delta-kick with amplitude
$(1-\e)$ to the field $E_x$, and a delta-kick with amplitude $\e$ to the field $E_y$, and
similarly for neurons $y$.
We will assume that both populations have equal number
of neurons $N$. The equations of the model thus read
\begin{gather}
\dot{x}_k = a_x - x_k + g_x((1 - \varepsilon)X + \varepsilon Y)\;,
\label{eq_oscX}\\
\dot{y}_k = a_y - y_k + g_y((1 - \varepsilon)Y + \varepsilon X)\;,
\label{eq_oscY}\\
\ddot{X}+2\alpha_x\dot{X}+\alpha_x^2 X=
\displaystyle \frac{\alpha_x^2}{N} \sum_{k,n_x}\delta(t - t_{k,n_x})\;,
\label{eq_fieldX}\\
\ddot{Y}+2\alpha_y\dot{Y}+\alpha_y^2 Y=
\displaystyle \frac{\alpha_y^2}{N} \sum_{k,n_y}\delta(t - t_{k,n_y})\;.
\label{eq_fieldY}
\end{gather}
Integration of these equations can be performed semi-analytically
(cf. Ref.~\cite{Zillmer_etal-07}). Between the firings, equations for $X,Y$ are linear
and the solution can be written explicitely, substitution of these
solutions to Eqs.~(\ref{eq_oscX},\ref{eq_oscY}) allows also for
an analytic representation of $x_k(t),y_k(t)$. This results in
transcendent equations for determining the next firing time, which
is solved numerically using the Newton method.

\section{Cross-Frequency Locking}
\begin{figure}[h!]
		\centering
			\includegraphics[width=\columnwidth]{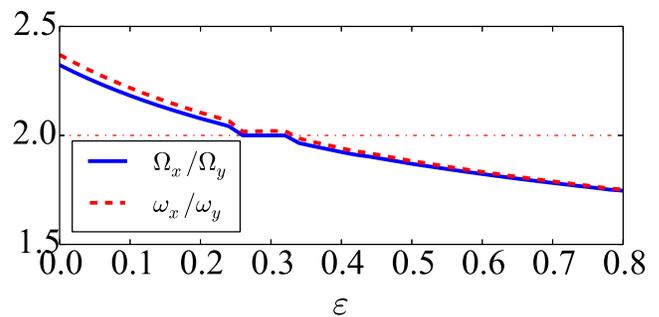}
			\caption{(Color online) Frequency ratios 
			$\Omega_x / \Omega_y$(solid line)  and $\omega_x / \omega_y$ (dashed line).
The 2:1 locking is observed for $\varepsilon \in [0.25; 0.33]$, 
			but in this case $\omega_x / \omega_y$ is not equal to 2. }
			\label{fig_21_freq_ratio}
		\end{figure}

\begin{figure}[h!]
\centering
\includegraphics[width=0.31\columnwidth]{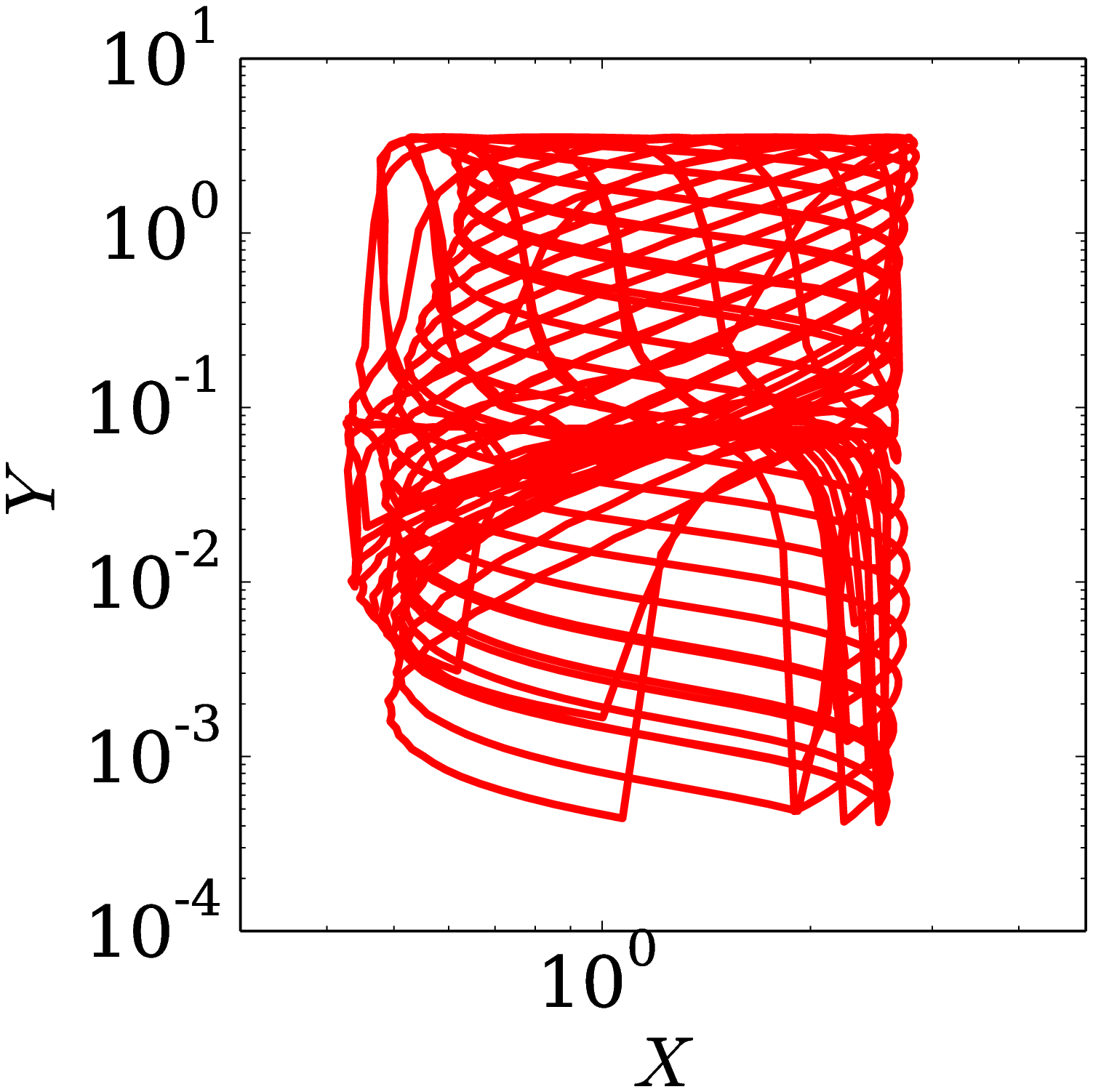}\hfill
\includegraphics[width=0.31\columnwidth]{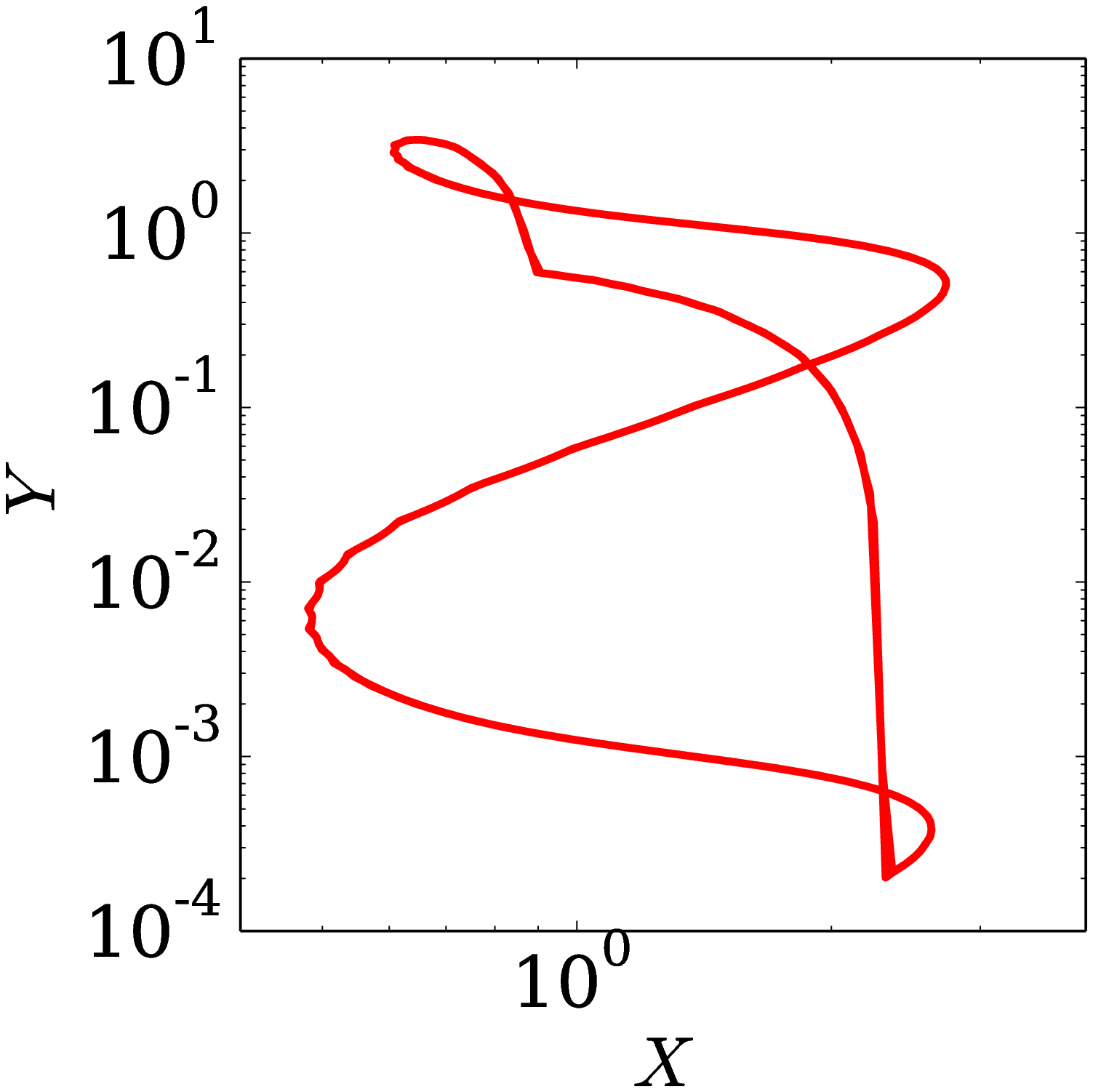}\hfill
\includegraphics[width=0.31\columnwidth]{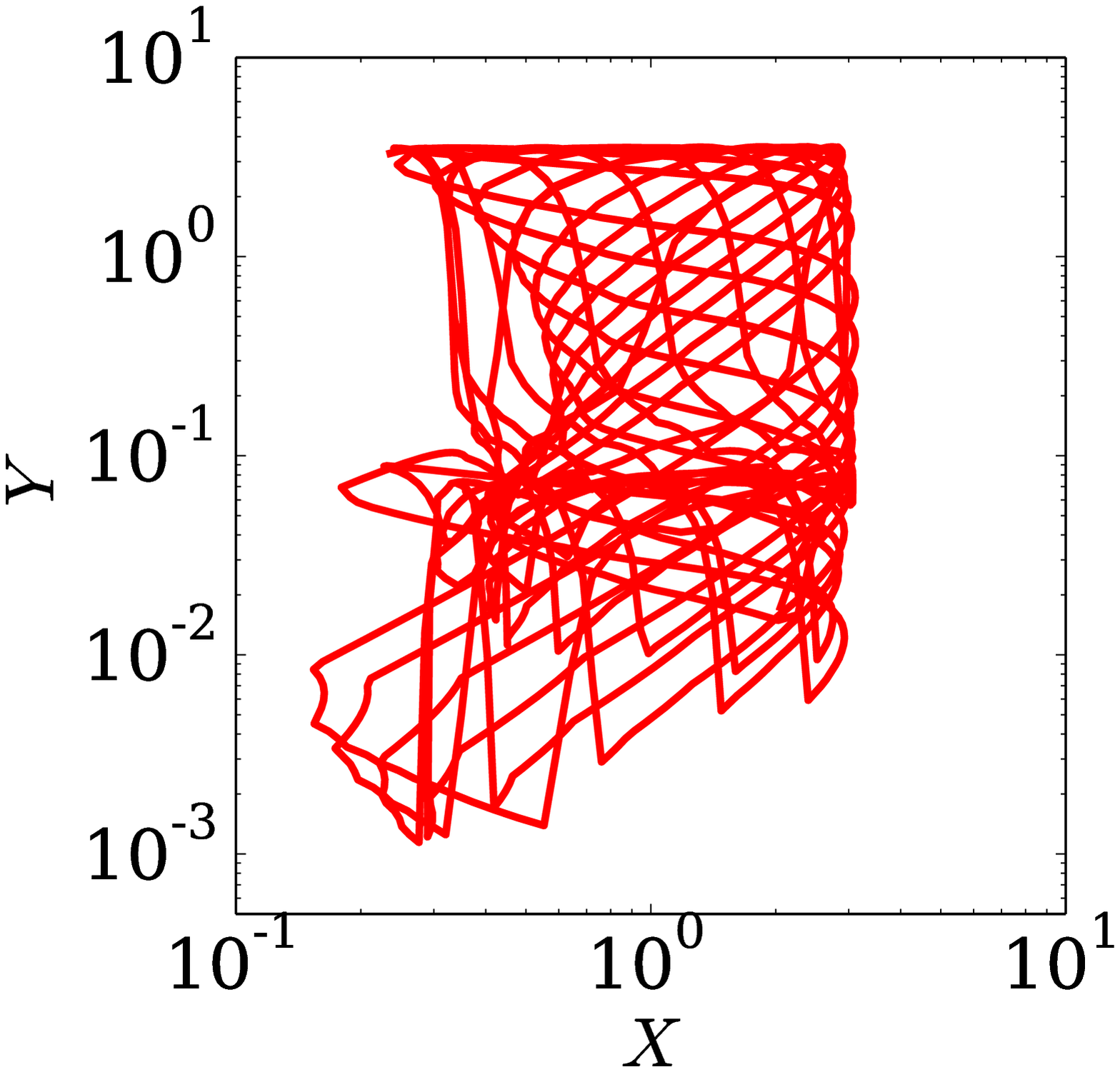}
\caption{Projections of the phase portraits on plane $(X,Y)$ for $\e=0.2$ (left panel),
$\e=0.3$ (middle panel) and $\e=0.4$ (right panel).}
\label{fig_21_Efields}
\end{figure}

In this section we focus on the effect of cross-frequency locking
in two interacting populations. As has been already mentioned, one
neural population demonstrates macroscopic mean-field oscillations
with frequency $\W$; two non-iteracting populations will have generally 
different macroscopic frequencies $\W_x,\W_y$. We demonstrate now,
that the interaction can lead to a rather nontrivial regime of
locking of these macroscopic oscillations. 
In Fig.~\ref{fig_21_freq_ratio} we report on the frequencies
of two neural populations for $a_x = 1.5$, $g_x = 0.35$, 
$a_y = 1.21$, $g_y = 0.09$, $\alpha_x = \alpha_y = 10$, $N=50$.
The figure shows the ratio of two macroscopic frequencies 
$\Omega_x / \Omega_y$ as a function of the coupling constant $\e$.
One clearly sees an interval of cross-frequency $2:1$ locking for
$\varepsilon \in [0.25; 0.33]$, here $\W_x=2\W_y$. Outside of this region
a macroscopic quasiperiodic regime with an irrational ratio $ \Omega_x / \Omega_y$
is observed as illustrated by the phase portraits projections on plane
$(X,Y)$ in Fig.~\ref{fig_21_Efields}. One clearly distinguishes the 
locked state at $\e=0.3$
from the quasiperiodic states at $\e=0.2$ and $\e=0.4$ on
these Lissagous-type curves.


\begin{figure}[h!]
\centering
\psfrag{k}[cc][cc]{text}
\includegraphics[width=0.75\columnwidth]{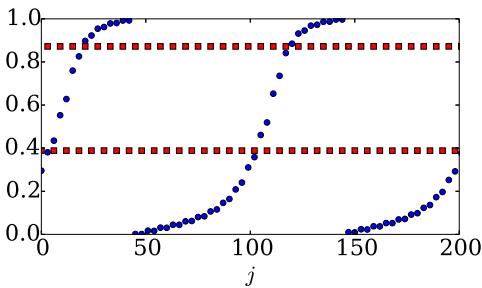}
\caption{(Color online) Stroboscopic observations of the states
of two oscillators in two populations, taken at $X(t^*_j)=1.5$, $\dot X>0$. 
Blue circles: values $x(t^*_j)$, red squares: values $y(t^*_j)$. Not all points are 
depicted for better visibility.}
\label{fig_21_xyfields}
\end{figure}

The frequency locking of the mean fields $X,Y$ does not mean that the
individual neurons in two populations are also mutually locked. We 
illustrate the dynamics of individual neurons in Fig.~\ref{fig_21_xyfields}.
Here we plot stroboscopic observations of one neuron from the population,
at moments of time when $X=1.5$ and $\dot{X}>0$. One can easily
compare the locking properties of two populations.
In population $x$ the states of the neuron are different, what indicates
that its firing frequency $\w_x$ is incommensurate with the frequency
$\W_x$ of the
mean field $X(t)$. In contradistinction to this,
 in population $y$ all the states are the same (they take two values because, due to 2:1 locking,
the mean field $X$ demonstrates two oscillations within one oscillation of the mean field $Y$)  
what means
that neurons in this population are locked by the mean field $Y(t)$ and fire
with the same period as the period of $Y(t)$, i.e. $\W_y=\w_y$. 
This property explains the behavior
of the ration of firing rates of neurons plotted in Fig.~\ref{fig_21_freq_ratio}:
in the whole region where one observes locking 2:1 of the mean fields $X,Y$,
the ratio $\w_x/\w_y$ of the firing rates of the neurons remains also a constant,
although not equal to $2$, but slightly exceeds this value due to a small difference
between $\w_x$ and $\W_x$.

\section{Marginal Chimera State}
\begin{figure}[h!]
\centering
\includegraphics[width=\columnwidth]{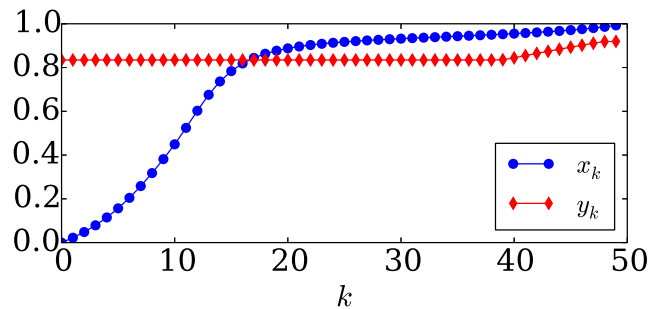}
\caption{(Color online)  Snapshots of states $x_k$, $y_k$ in the 2:1 locked state for $\e=0.3$.
While all values of $x_k$ are different, in population $y$ the neurons with $1\leq k\leq 39$
form a cluster.}
\label{fig_03_state}
\end{figure}

Locking of the neuron states in ensemble $y$ by the mean field acting 
on these neurons, at $\e=0.3$, allows one to expect that these neurons
form a fully synchronized cluster. Surprisingly, this is not the case. In 
Fig.~\ref{fig_03_state} we plot the states of all neurons in two populations
at a certain moment of time, in the 
locked regime $\e=0.3$. In population $x$ all the states are different, what
corresponds to the fact that they are in a quasiperiodic mode of partial synchronization.
Neurons in population $y$ are in a chimera state: a part of them build a fully
synchronized cluster $y_1=y_2=\ldots=y_m$, while all the neurons $y_k$, $k>m$,
are distributed in some range. We have checked that this property is not an artifact, by
observing it for different population sizes and for very long evolution times (up to $10^6$).

At first glance, a chimera state in a population of identical
oscillators described by first-order equations
(like Eqs.~(\ref{eq_oscX},\ref{eq_oscY}) of our model) is impossible. Indeed, 
in the case of a periodic forcing by mean fields
$(X(t),Y(t))$, Eq.~\eqref{eq_oscY} for a neuron in population $y$
reduces to a one-dimensional circle map, if a stroboscopic map
is constructed from this first-order equation. This map is the same for all neurons in
population $y$. According to general theory of one-dimensional  circle maps,
all neurons have the same frequency (because the rotation number of one-dimensional
maps does not depend on initial conditions). Moreover, for general one-dimensional maps
one has a dichotomy~\cite{Hasselblatt-Katok-03}: (i) 
either there is an equal number of unstable and stable periodic orbits, the 
latter attract all the points of the circle except for those lying exactly on
unstable orbits; (ii) or the dynamics
is quasiperiodic and reduces according to Denjoy's theorem to a shift on the circle, here
all initially different states remain different. This dichotomy allows for quasiperiodic (partial synchronization)
and fully synchronized regimes, but seemingly excludes chimera states.

\begin{figure}[h!]
\centering
(a)\includegraphics[width=0.7\columnwidth]{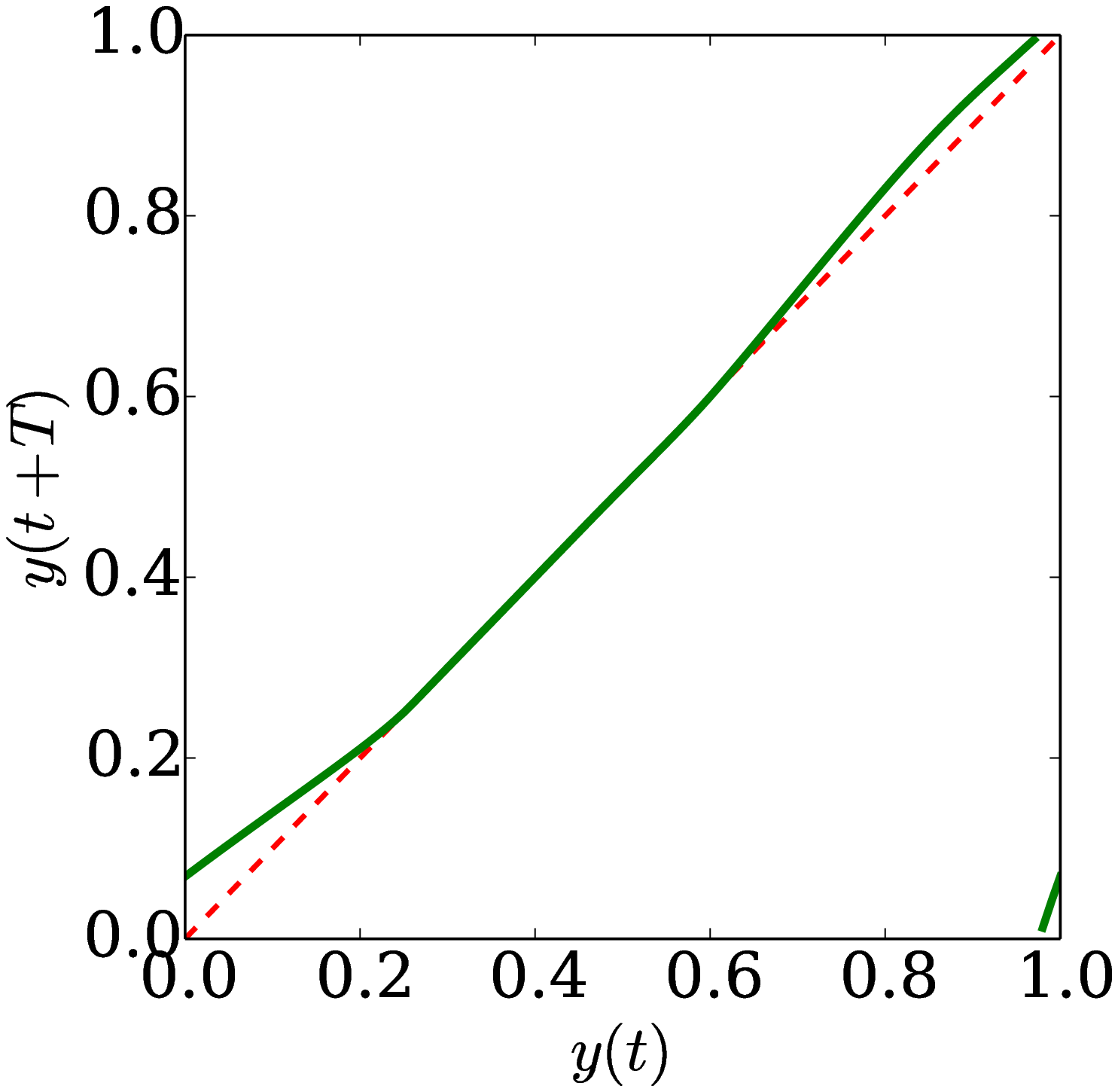}
(b)\includegraphics[width=0.7\columnwidth]{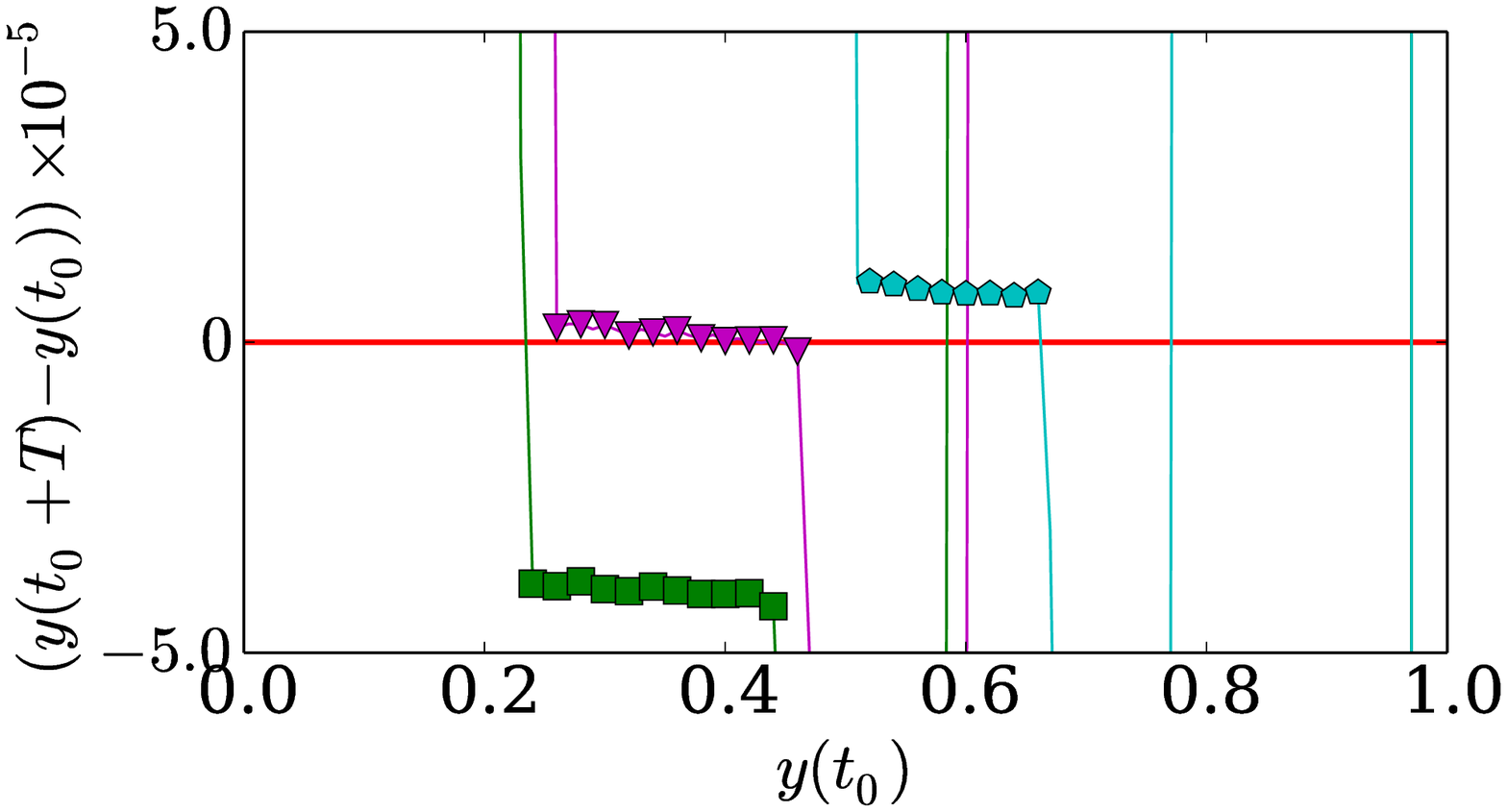}
\caption{(Color online) (a) The stroboscopic map $y(t)\to y(t+T)$ for neurons in population
$y$ in the 2:1 locked regime for $\e=0.3$. For the chosen phase of the mean field, in the
region $0.25\lesssim y\lesssim 0.45$ the map is nearly identity. This region is resolved in
panel (b), where we plot values $y(t_0+T)-y(t_0)$ vs $y(t_0)$ for different choices of $t_0$.
Small vertical shifts are due to a tiny degree of qusiperiodicity in the fields $X(t),Y(t)$. }
\label{fig_03_map}
\end{figure}

In the dynamics of our two coupled populations we see, that the population $y$ violates the dichotomy
above. To clarify this point, we constructed a stroboscopic map $y(t_0)\to y(t_0+T)$, 
where $T$ is period of the field
$Y(t)$ (Fig.~\ref{fig_03_map}) and $0\leq y(t_0)<1$. 
One can see that this map is not of general smooth type, as it has 
an interval  (we call it marginal domain) where $y(t_0+T)\approx y(t_0)$ with 
high accuracy. This can be seen in panel (b) of Fig.~\ref{fig_03_map},
where the marginal domain is enlarged. Here the deviations $y(t+T)-y(t)$ are of order $10^{-5}$. 
Moreover, these deviations fluctuate in sign, if
the stroboscopic map is built at different phases $t_0$ of the mean fields $X(t_0),Y(t_0)$. 
These fluctuations are
due to the fact, that the fields $Y(t)$ is, strictly speaking, 
not exactly periodic. Indeed, due to quasiperiodicity
of the population $x$, the fields $X(t),Y(t)$ are quasiperiodic, although deviations from the pure periodicity
are extremely small and are not seen in Fig.~\ref{fig_21_Efields}. This quasiperiodicity can, e.g., 
be seen if one  at a given value of $Y$ plots values of $\dot{Y}$, these are spread in a small interval. Due to this
small spreading, the  deviations from identity map fluctuate in the marginal domain.

Existence of the marginal domain, together with a complementary interval where  $y(t_0+T)\neq y(t_0)$, explains the
observed chimera state: those neurons which have initial conditions in the non-marginal domain, are attracted to one
state and form a cluster, while those in the marginal domain remain scattered and form a ``cloud''.  Noteworthy,
the described non-general properties of the dynamics are not pre-built to the system, but appear in a 
self-consistent manner, because the fields $X,Y$ are composed from the contributions from individual neurons.
Quite unexpectedly, these mean fields are self-organized in such a manner, that one population is purely
quasiperiodic, while another one combines properties of stable and marginal dynamics that results in a chimera.

\section{Conclusion}

In this paper we considered two populations of integrate-and-fire oscillators having 
definitely different frequencies of the generated mean fields. First we showed, that
due to mutual coupling, a 2:1 locking of the mean fields can be observed, without
synchronization between the individual neurons. Noteowrthy, neurons
in two populations behave differently in the locked state: 
while in one population they are not synchronized by the mean field and have a
different
frequency, in another population the period of firings is the same as the basic period 
of the mean field. However, this synchronous state is rather nontrivial, and this
is our second main result:
 identical neurons in the synchronous population build a chimera: one part
of them forms an identical cluster, while other oscillators do not join this cluster and
remain scattered. We explain this regime as a self-sustained marginal dynamics of the
driven neurons: the corresponding stroboscopic one-dimensional map has a domain
where this map is practically an identity. This marginality is possibly the only
way to achieve a bistability in a one-dimensional map, as the period here must be 
independent of initial conditions. This is another peculiarity of the marginal chimera:
in other cases where chimera has been observed in 
identical units, the frequencies of a cluster and
a scattered populations were 
different~\cite{Schmidt_etal-14,*Sethia-Sen-14,*Yeldesbay-Pikovsky-Rosenblum-14}.

We acknowledge discussions with S. Olmi, A. Politi, and A. Torcini.
M. B. was supported by ITN COSMOS (funded by 
the European Union’s Horizon 2020 research and innovation
programme under the Marie Sklodowska-Curie grant agreement No 642563).
This work (Sections I-III) was  
supported by the Russian Science Foundation 
(Project No. 14-12-00811) and by the grant of  
The Ministry of education and science of the Russian 
Federation (the agreement of August 
27, 2013 № 02.В.49.21.0003) (Sections IV-V).
  
%

\end{document}